\documentclass[aps,showpacs,twocolumn]{revtex4}
\usepackage{epsfig}
\usepackage{amsfonts}
\usepackage[T1]{fontenc}

\newcommand{\beq}{\begin{equation}}
\newcommand{\eeq}{\end{equation}}
\newcommand{\bea}{\begin{eqnarray}}
\newcommand{\eea}{\end{eqnarray}}

\newcommand{\al}{\alpha}
\newcommand{\s}{\sigma}

\newcommand{\be}{\beta}

\newcommand{\ra}{\rangle}
\newcommand{\la}{\langle}

\begin{document}

\title{Nuclear spin ferromagnetic phase transition in an interacting $2D$ electron gas}
\author{Pascal Simon$^{1,2}$ and Daniel Loss$^1$}
\affiliation{$^{1}$ Department of Physics and Astronomy, University of
Basel, Klingelbergstrasse 82, CH-4056 Basel, Switzerland}
\affiliation{$^{2}$ Laboratoire de Physique et Mod\'elisation des  Milieux
    Condens\'es, CNRS and Universit\'e Joseph Fourier, BP 166, 38042
Grenoble, France}

\date{\today}
\begin{abstract}
Electrons in a two-dimensional semiconducting heterostructure interact with nuclear spins
via the hyperfine interaction.
Using a  a Kondo lattice formulation of the electron-nuclear spin interaction, 
we show that the nuclear spin system within an interacting two-dimensional electron gas undergoes a ferromagnetic 
phase transition at finite temperatures.
We find that electron-electron interactions and non-Fermi liquid behavior substantially 
enhance the nuclear spin Curie temperature into the $mK$ range with decreasing electron density. 

\end{abstract}

% 71.10.Ca Electron gas, Fermi gas
% 71.27.+a Strongly correlated electron systems, heavy fermions
% 71.45.Gm 	Exchange, correlation, dielectric and magnetic response functions, plasmons
% 71.70.Gm 	Exchange interactions
% 75.30.Mb 	Valence fluctuation, Kondo lattice, and heavy-fermion phenomena
%71.10.Ay    Fermi-liquid theory and other phenomenological models

\maketitle

%{\em Introduction}.
The use of the electron spin  as a qubit for quantum computing
relies on the ability to coherently control  single electron spins in semiconductor 
quantum dots \cite{loss:1998}. 
%This is the key stone towards scalable spin-based 
%quantum information storing and quantum computing. 
%In the last decade,
%all the necessary requirements for spin-based quantum gates have been realized experimentally
%in GaAs semiconductors, ranging from the coherent exchange of two electron spins in a double dot \cite{petta:2005}
%to the coherent control and observation of Rabi oscillations
%of  single-electron spins \cite{koppens:2006}. 
Over the last years much progress has been made for dots in
GaAs semiconductors, where single spin lifetimes have been measured to range
far into the $\mathrm{ms}$-range  \cite{kroutvar:2004a,elzerman:2004,amasha:2006a}, and where
coherent manipulation of single- and two-spin 
states was successfully implemented \cite{petta:2005, koppens:2006}.
Still, a major obstacle to further progress is the comparatively short spin decoherence
time in these materials, ranging  from  $100\,\mathrm{ns}$ in bulk \cite{kikkawa:1998}  to  $\mu s$ in dots\cite{petta:2005}. 
%These achievements have been made 
%possible because electron spins
% in GaAs quantum dots are  
%relatively isolated from their surrounding environment and therefore long lived quantities
%robust against decoherence. Indeed, longitudinal
%relaxation times in these systems have been measured to be in the $\mathrm{ms}$ range
%%
%%of order $\mathrm{ms}$
%%in a magnetic field of $8\,\mathrm{T}$
% \cite{kroutvar:2004a,elzerman:2004,amasha:2006a}.
%Furthermore, a lower bound on the spin coherence time exceeding 1$\mu s$ has been recently established using spin-echo techniques \cite{petta:2005}.
%%the transverse dephasing time $T_{2}^{*}$ in GaAs quantum dots
%%for an ensemble of electron spins 
%%has been measured to be typical larger than  $100\,\mathrm{ns}$ \cite{kikkawa:1998}.
The main source of decoherence for a single electron spin confined
to a GaAs dot is coming from the contact hyperfine
interaction with the surrounding nuclear spins \cite{burkard:1999a,khaetskii:2003,coish:2004a}.
% \cite{burkard:1999a}.
Several ways to overcome this problem have been proposed
such as
spin echo techniques \cite{coish:2004a,petta:2005}, projection of the nuclear spin state \cite{coish:2004a} or polarization of the nuclear spins \cite{burkard:1999a,khaetskii:2003,coish:2004a,imamoglu:2003}. However, in order to 
extend the spin decay time  by one order of magnitude,
%through polarization of the nuclear spins, 
a polarization of  above 99\% is required \cite{coish:2004a}, which is still 
 far away from the $60\%$ so far reached  in quantum dots 
 via optical pumping  \cite{bracker:2005a}.
%
%the best result so far reached in quantum dots with optical pumping which is around $60\%$ \cite{bracker:2005a}.
% 
One way to overcome this problem would be that nuclear spins become fully 
polarized at low enough temperatures,
without any external magnetic field or optical pumping. This is the case if the
nuclear spins  undergo 
a ferromagnetic phase transition at a finite Curie temperature $T_c$.
%Actually, it turns out that 
Quite remarkably, the possibility of such a nuclear-spin phase transition to occur in
a metal
%the question 
%about the value of the nuclear spin Curie temperature 
%in an metal was raised more
was studied more 
than sixty years ago by Fr\"ohlich and Nabarro (FN) \cite{FN:1940}.
Using a Weiss mean field argument they showed that the
Curie temperature $T_c$ of nuclear spins in a three dimensional ($3D$) metal becomes
\beq \label{eq:FN}
k_BT_c\sim \frac{A^2}{8E_F}\, ,
\eeq
where $A$ denotes the hyperfine coupling strength between the nuclear and  electron spin
 and $E_F$  the Fermi energy.
%Assuming the electron is in an s-state, the hyperfine Hamiltonian reads
%$H_{HF}=A \vec I\cdot \vec \s$ where $\vec I$ is the nulcear spin and $\s$ the spin of the electron.
For a typical metal, $T_c$  is of the order of micro-Kelvin or less. However, for a 
two-dimensional electron gas (2DEG) in GaAs semiconductors, Eq. (\ref{eq:FN})
 would predict
nuclear ferromagnetism with $T_c \sim 1 {\mathrm mK}$, which is surprisingly high 
%and almost comparable to low temperature nowadays reached in a laboratory to 
%study properties of semiconducting heterostructures. 

However, the direct use of Eq. (\ref{eq:FN}), which was derived for a bulk metal, to 
a 2DEG in a semiconductor  is  very problematic. 
%One particular reason is that, according to the 
%well-known Mermin-Wagner theorem (MWT) \cite{MW}, a phase transition
%at finite temperatures requires much more stringent conditions
%in 2D (and 1D) than in 3D.
The purpose of this letter, therefore, is to reconsider this issue for a 2DEG and
to estimate the nuclear spin Curie temperature. Our analysis below will be based
on the Kondo lattice
model \cite{sigrist:1997}, where we integrate out the electron degrees of freedom to derive an effective
spin Hamiltonian whose exchange is given in terms of the static electronic  spin susceptibility $\chi_{s}(q)$. 
Using a spin-wave analysis, 
we will show that the electron-electron (e-e) interactions in the 2DEG and the induced
 non-Fermi liquid behavior in $\chi_{s}(q)$ \cite{belitz:1997,maslov:2003,maslov:2006,efetov:2006} ultimately enables a ferromagnetic phase transition of the nuclear spins. For sufficiently
 strong interactions and/or low electronic densities (with the dimensionless interaction parameter 
$r_{s} \sim 5-10$) the Curie temperature
 can be pushed  into the milli-Kelvin regime, and thus, the phase transition should become accessible
 experimentally.
 
{\em Model Hamiltonian}.
In order to study an interacting 2DEG coupled to nuclear spins within the 2DEG, we adopt a tight-binding representation in which
each lattice site contains a single nuclear spin and electrons can
hop between  neighboring sites.  The Hamiltonian describing such
a system  reads
\beq\label{eq:kl}
H=H_0+\frac{1}{2}\sum\limits_{j=1}^N A_j c^\dag_{j\al}\vec \sigma_{\al\be} c_{j\be}\cdot \vec I_j=H_0+H_n,
\eeq
where 
$H_0$ denotes the conduction electron Hamiltonian and $H_n$ the electron-nuclear spin hyperfine 
interaction. 
$H_0$ can be rather general and 
includes e-e interactions. 
In Eq. (\ref{eq:kl}), $c^\dag_{j\al}$ creates an electron at the lattice site 
$\vec r_j$ with spin $\al$ and $\vec \sigma$ represent the Pauli matrices.
We have also introduced $\vec I_j$ the
nuclear spin located at the lattice site 
$\vec r_j$, and $A_j$ the hyperfine coupling constants between the electron
and nuclear spins at site $\vec  r_j$. The electron spin operator is defined by $\vec S_j= \frac{1}{2}c^\dag_{j\al}\vec \sigma_{\al\be}c_{j\be}$.
$N$ denotes the total number of
sites on the $2D$ lattice. In our formulation, the nuclear spin density is $n_s=a^{-2}$ where $a$ is the lattice spacing. From here on, we assume $A_j=A>0$
which means we assume the hyperfine interaction to be the same for all atoms
that constitute the heterostructures (typically Ga and As).
We also neglect direct dipolar interactions between the nuclear spins
which is in general smaller than the indirect interaction as we will see.
This amounts to assume that the dipolar interaction energy scale $E_{dip}$ is among the smallest 
one
and particularly that $k_BT\gg E_{dip}$, where $T$ is the temperature.
This assumption is crucial since it allows us to focus on the nuclear spins which are
within the $2D$ electron gas thickness (in growth direction) and justifies our $2D$ description \cite{note:3d}.

The general Hamiltonian in Eq. (\ref{eq:kl}) is the well-known $2D$ Kondo lattice Hamiltonian (KLH),
though $H_0$ contains also e-e interactions.
The regime we are interested in corresponds to the weak Kondo coupling regime 
in the sense that $A\ll E_F$,
where $E_F$ is the Fermi energy.
The KLH has been introduced to describe various physical properties of
heavy-fermion materials \cite{lee:1986,sigrist:1997}, and more
recently also of ferromagnetic semiconductors \cite{fs}.
%The KLH has also been 
%proposed recently to describe properties
%of  ferromagnetic semiconductors \cite{fs}. 
%In a 2DEG, the electron density $n_e$ is much lower than $n_s$. 
%In this low density regime, 
%the ground state of the magnetic system has been shown to be 
%ordered ferromagnetically in  $3D$ 
%using various treatments that go beyond mean field theory and notably include spin wave modes \cite{sigrist:1997}.
%For the oversimplified case of a single electron interacting with a lattice of spin impurities, 
%this statement can be even proved  exactly \cite{sigrist:1991}. 

Before turning to the extended system let us briefly consider the special case
of a single electron confined to a quantum dot which interacts typically with $10^6$
nuclear spins \cite{khaetskii:2003,coish:2004a}.
%We note in passing that the single electron case 
%is relevant to describe the interaction between a single electron trapped in a quantum dot 
%and the neighboring nuclear spins. 
This case can be described by the above KLH by allowing  in $H_0$ for a confinement potential for the dot, which 
provides the largest energy scale. 
%It is indeed enough to include a strong confining potential in $H_0$. If we assume 
%the confining potential has the highest energy scale, 
Indeed, we can then project
$H_n$ into the ground state of $H_0$,
%(which now includes the single electron kinetic energy plus the confining
%potential), 
and the hyperfine Hamiltonian then takes the known central spin form 
$H=\sum_i \widetilde A_i\vec S_e\cdot \vec I_i$ \cite{khaetskii:2003,coish:2004a}, where $\vec S_e$ is the single electron spin, and
$\widetilde A_i=A|\psi(\vec r_i)|^2$ the non-uniform coupling constant with
$\psi(\vec r_i)$ the electronic ground state wave function at site $\vec r_i$. 
The reformulation of the central spin problem in terms of the KLH should be
particularly useful for numerical evaluations.
%In this sense, our KLH description of the hyperfine interaction is much more general
%and also encompasses the quantum dot geometry provided a confining potential is added.

%Let us continue now with the extended case. 
%In order to make progress 
%with the Hamiltonian in Eq. (\ref{eq:kl}),
To continue with the general case,
 it is convenient to go to  Fourier space and rewrite $H_n$ in Eq. (\ref{eq:kl}) as
%$H_n$ in Fourier space as
%\beq\label{eq:kl1}
$H_n=\frac{A}{N}\sum_{\vec q} \vec S_{\vec q}\cdot\vec I_{\vec q}$,
%\eeq
where
$\vec I_{\vec q}=\sum_j e^{-i \vec q\cdot\vec r_j}\vec I_j$ 
 %$\vec S_{\vec q}=\sum_j e^{-i \vec q\cdot\vec r_j}\vec S_j$, 
 is the Fourier transform of $\vec I_j$, and similarly for $\vec S_{\vec q}$.  Since $A$ is a small energy scale in our case, we may perform a Schrieffer-Wolff (SW) transformation
in order to eliminate terms linear in $A$, and thereby  integrate out the electronic
degrees of freedom.
Keeping the lowest order terms in $A^2$ of the SW transformation, we are left with an effective Hamiltonian
$
H_{eff}=H_0-\frac{1}{2}[S,[S,H_0]]$.
$S$ is defined by $H_n+[S,H_0]=0$, which is solved as $S=L_0^{-1}H_n$
where $L_0$ is the Liouvillian. Let us define $U=\frac{1}{2}[S,[S,H_0]]$
which can be rewritten as $U=\frac{1}{2}[L_0^{-1}H_n,H_n]$.
Using an integral representation for  $L_0$, one obtains
$
U=-\frac{i}{2}\int_0^{\infty}dt e^{-\eta t} [H_n(t),H_n],
$
%$
%U=-\frac{i}{2}\int\limits_0^{\infty}dt e^{-\eta t} [H_n(t),H_n],
%$
where $\eta \to 0^+$ ensures convergence.
We next take the equilibrium expectation value over electronic
degrees of freedom, denoted by $\la\dots\ra$.  The only assumptions we make are 
$\la S_i^x\ra=\la S_i^y\ra=0$,
and translational invariance in the 2DEG. We then get
% Note that $\la\dots\ra$ means here trace over electronic degrees of freedom only.
%This allows us to bring $\la U\ra$ in a the following form,
\beq\label{eq:ueff}
\la U\ra=\frac{A^2}{8n_s}\sum\limits_{\vec q} I_{\vec q}^\al~
\chi_{\al \be}( q) ~I^{\be}_{-\vec q}~,
\eeq
where
$
\chi_{\al\be}( q)=-i\int_0^\infty dt~ e^{-\eta t}\la[ S_{\vec q}^\al,S_{-\vec q}^\beta]\ra, $
and where
summation over the spin components $\al,\beta=x,y,z$ is implied.
%
%$
%\chi_{\al\be}( q,\omega)=-i\int\limits_0^\infty dt~ e^{-i \omega t-\eta t}\la[ S_{\vec q}^\al,S_{-\vec q}^\beta]\ra.
%$
%is the electronic spin susceptibility. 
If we also assume $\la S^z_i\ra=0$, then 
$\chi_{\al \be}(q)=\delta_{\al \beta} \chi_{s}(q)$, where $\chi_{s}(q)$ 
is the electronic  spin susceptibility in the static limit.
We stress that Eq. (\ref{eq:ueff}) is rather general and requires only weak assumptions
on $H_0$. 
%To the best of our knowledge, this has not been noticed before. 
In real space we have
%\beq\label{eq:hreal}
$\la U\ra=-\frac{1}{2}\sum_{\vec r,\vec r'} J_{\vec r-\vec r'}^{\al\be}
 I_{\vec r}^\al I_{\vec r'}^\be$,
%\eeq
where 
%$ J_{\vec r}^{\al\be}=-\frac{A^2}{4n_s}\chi_{\al\be}(\vec r)$. 
$ J_{\vec r}^{\al\be}=-({A^2}/{4n_s})\chi_{\al\be}(\vec r)$ is the effective exchange coupling.
The nuclear spins $\vec {I}_{\vec r}$
are therefore interacting with each other, this interaction being mediated by the conduction electrons.
This is nothing but the standard  Ruderman-Kittel-Kasuya-Yosida (RKKY) interaction, which, as we shall
see, can be substantially modified by e-e interactions compared to the free electron case.

Let us first analyze the case of non-interacting electrons. In this case,
$\chi_{s}$ coincides with the usual density-density response (Lindhard) function $\chi_{0}$ \cite{GV}.
%\beq
%\chi_{0}(q)=\frac{1}{Na^2}\sum\limits_{\vec k,\s}\frac{n_{\vec k,\s}-n_{\vec k+\vec q,\s}}
%{\epsilon_{\vec k,\s}-\epsilon_{\vec k+\vec q,\s}+i\hbar\eta},
%\eeq
%where $n_{\vec k}$ is the electronic number operator and $\eps_{\vec k}$ the dispersion relation.
We first perform a mean field analysis.
The Weiss mean field theory predicts a Curie temperature
\beq\label{eq:mf}
T_c=-\frac{I(I+1)}{3k_B}\frac{A^2}{4n_s}\chi_0(q=0),
\eeq
where $I$ is the nuclear spin value.
In $2D$, $\chi_0(q=0)=-N_e= -m^*/\pi$, where 
$N_e=n_e/E_F$ is the electronic 
density of states, and 
$m^*$ is the effective electron mass in a 2DEG (we set $\hbar=1$).  
%The Weiss mean field result predicts a dependence of $T_c^{CW}$ with the ratio $n_e/n_s$. 
For a $3D$ bulk metal with one conduction electron per nucleus, the ratio $n_e/n_s\sim 1$ and
we recover the result in Eq. (\ref{eq:FN}) derived more than sixty years ago by Fr\"ohlich and Nabarro \cite{FN:1940}.
For a $2D$ metal, 
%(see below for the role of the Mermin-Wagner theorem)
the Weiss mean field theory predicts $k_BT_c=I(I+1)A^2/12E_F$.  
For a $2D$ semiconductor, however, the ratio $n_e/n_s$
is much smaller than $1$. With typical values for GaAs heterostructures,
$I=3/2$, $A\sim 90~\mu eV$ and $a\sim$ 2\AA \cite{coish:2004a}, we estimate $T_c\sim 1~\mu K$, which is very low. 
(For such low $T_{c}$'s, ignoring nuclear dipole-dipole interactions from the start would not be valid.) However, this estimate is just based on the simplest mean field theory and, moreover, does not include the effect 
of e-e interactions. 
%It still predicts a finite $T_c$ under which the nuclear spins
%order ferromagnetically. 

We shall now go beyond above mean field approximation. For this we assume  that the ordering 
(if it takes place) leads to a ferromagnetic phase where the collective low-energy excitations
are given by spin waves. Then,  we define the Curie temperature  $T_{c}$ as the temperature
at which the magnetic order is destroyed by those spin waves. This procedure is equivalent to the
Tyablikov decoupling scheme \cite{Tyablikov}.
%The low excitations of a quantum ferromagnet are spin waves which are collective excitations. 
%Another way of determining the Curie temperature  is to analyze at which
%temperature $T_c$ the spin wave analysis breaks down. 
The dispersion relation of  the spin wave (or magnon) reads
\beq
 \omega_q=I(J_0-J_{q})=I \frac{A^2}{4}a^2(\chi_{s}({q})-\chi_{s}(0)),
\eeq
where $J_{q}$ is the Fourier transform of $J_{\vec r}$.
The magnetization $m$ per site  at finite $T$ is $m(T)=I-\frac{1}{N}\sum_{\vec q} n_q$,
where 
%$n^m_{q}=(e^{\beta\omega_q}-1)^{-1}$ 
$n_{q}=(e^{\omega_q/k_{B}T_{c}}-1)^{-1}$
is the magnon occupation number.
%and $\beta=1/k_BT$. 
The Curie temperature $T_c$ follows then from the vanishing of the magnetization,
 i.e. $m(T_c)=0$,  which, in the continuum limit, becomes
\beq\label{eq:tc}
1=\frac{a^2}{I}\int \frac{d\vec q}{(2\pi)^2} \frac{1}{e^{ \omega_q/k_{B}T_{c}}-1}.
\eeq
%If one assumes $\hbar \omega_q\ll{k_B T_c}$ one may expand the denominator and the Curie temperature reads
%\beq
%k_BT_c^{SW}=\frac{A^2I^2}{4}\frac{1}{\int\limits_0^\infty 
%\frac{d\vec q}{(2\pi)^2}(\chi_{s}(q)-\chi_{s}(0))^{-1} }
%\eeq
%This expression has been derived in Ref. \cite{sigrist:1997} under some Tyablikov-decoupling 
% approximation scheme
%and starting with non-interacting electrons. Nevertheless, the approximation $\hbar \omega_q\ll{k_B T_c}$ is not really justified and it seems safer to determine $T_c^{SW}$ from the general equation 
%(\ref{eq:tc}) which is also valid for interacting electrons.
%We note that the integral in Eq. (\ref{eq:tc}) is dominated by small $q$-values. 
%One may therefore expand
%the exponential and determine directly $T_c$ as an integral formula \cite{sigrist:1997}; however,
%this expansion is not necessary. as we will see. 
For non-interacting electrons in $2D$, $\chi_{s}(q)-\chi_{s}(0)=0$ for $q< 2k_F$ \cite{GV},
where $k_F$ is the Fermi wave vector. The spin wave analysis therefore predicts $T_c=0$, 
in agreement with a recent conjecture extending the Mermin-Wagner theorem
for RKKY interactions in a non-interacting 2D system \cite{bruno:2001}.
%This result could have been anticipated from the Mermin-Wagner theorem (MWT) \cite{MW}. Indeed,
%this theorem states that there is no phase transition at finite
%$T$ for spin systems in $1D$ and $2D$ with isotropic Heisenberg  interactions provided
%that the interactions are short range enough more precisely  that 
%$\sum_{\vec r} r^2 J_{\vec r}<\infty$, which  is equivalent in $q$-space to  $\lim_{q\to 0}\vec\nabla_q^2 J_{q}<\infty$.
%$\lim\limits_{q\to 0}\vec\nabla_q^2 J_{q}<\infty$
%For $2D$ non-interacting electrons, 
%the nuclear spins interact with a long range interaction which decays
%in $1/r^2$ in $2D$ but with an oscillatory character. It results that 
%$J_{ q}-J_{q=0}=0$ for $q< 2k_F$ and the MWT applies, enforcing $T_c=0$. Note that the Weiss mean field result for 2D free electrons
% violates the MWT, which, of course, is not so surprising given the reduced dimensions.
% since 
%the mean field theory does not take into account properly the dimensionality
%of the system.
%For interacting electrons, however, the long range decay of the RKKY interactions can be altered
%substantially and no conlcusion can be {\it a priori} drawn from the MWT or its extensions.
%which then prevents the application of the MWT, as we shall see next.
 
 The study of thermodynamic quantities in {\em interacting}  electron
 liquids especially in 2D
 has attracted quite some interest recently  with the goal to find deviations from the standard Landau-Fermi liquid behavior, such as non-analytic dependences on the wave vector  \cite{belitz:1997,maslov:2003,maslov:2006,efetov:2006}.
 %Incorporating e-e interactions in the calculations of thermodynamic quantities has been %at the heart of condensed matter theory over the last fifty years. In particular, 
%The study of non-analytic behavior of thermodynamical quantities in electron liquids has attracted recent interest 
%especially in $2D$ \cite{belitz:1997,maslov:2003,maslov:2006,efetov:2006}. 
In particular, it was found \cite{maslov:2003} that the static non-uniform spin susceptibility $\chi_s(q)$ depends {\em linearly } on the wave vector $q=|\vec q|$ for $q\to 0$ in 2D (while it is $q^2$
in 3D). This non-analyticity arises from the long-range correlation between quasiparticles mediated by virtual particle-hole pairs.
Since the integral in Eq. (\ref{eq:tc}) is dominated by the low $q$-behavior,
one may replace $\omega_q$ by its low-$q$ limit which
turns out to be  linear in $q$ (see
below) \cite{dispersion}.
%\bibitem{dispersion}
%Such a linear spin wave dispersion  is usually associated
%with antiferromagnets while one would expect a quadratic dispersion for ferromagnetically ordered
%states like considered here. 
%This somewhat unexpected linear dispersion comes purely from  e-e interactions. 
The integral in 
Eq. (\ref{eq:tc})
can  then
be  performed easily,
allowing us to express $T_c^{}$ in terms of the derivative of the spin sucseptibility,
\beq \label{eq:tcsw}
T_c^{}=\frac{A^2 I}{2k_B}\sqrt{ \frac{3I}{\pi n_s}}\left.
\frac{\partial \chi_s(q)}{\partial q}\right|_{q\to 0}.
\eeq  
For non-interacting electrons, $\delta\chi_s(q)=0$ at low $q$ and we recover $T_c^{}=0$,
in accordance with the MWT.

Let us include now e-e interactions. To calculate $\chi_s(q)$, we start from the Bethe-Salpeter (BS)
equation for the two-body scattering amplitude \cite{GV}. Solving the
 BS equation formally, we can derive an exact and closed expression for the spin susceptibility given by
\beq\label{eq:chis}
\chi_s(\bar q)=\frac{1}{L^{2D}}\sum\limits_{\bar p,\bar p'}\left( R(\bar q)\frac{1}{1-\Gamma^-_{ir}(\bar q)R(\bar q)}\right)_{\bar p\bar p'}\, ,
\eeq
where $L=\sqrt {Na^2}$ is the system length, $ (\Gamma_{ir}^-)_{\bar p\bar p'}(\bar q)$ the exact irreducible electron-hole
 scattering amplitude in the spin channel
%(see \cite{GV}), $R_{\bar p}(q)=\frac{}{ Na^2} 
(see \cite{GV}), $R_{\bar p}(\bar q)=
-2i G(\bar p+\bar q/2)G(\bar p-\bar q/2)$ is the electron-hole bubble where $G(\bar p)$ is the exact propagator and $\bar p\equiv (p_0,\vec p)$ is the  (D+1)-momentum with $p_0$ the frequency.
%(we have introduced a relativistic notation where $p_0$ denotes the frequency and $\vec p$ the $2D$ wave vector).
We have used a matrix notation in Eq. (\ref{eq:chis}) where the indices run over 
 $\bar p$ ($R$ is a diagonal matrix). 
Unfortunately, $\Gamma^-_{ir}$ cannot be calculated exactly and some drastic approximations are required.
The approximation we use consists in replacing the
exact irreducible electron-hole scattering amplitude
$(\Gamma^-_{ir})_{\bar p,\bar p'}$  by an averaged value calculated 
with respect to all possible values 
of $p$ and $p'$
near the Fermi surface, therefore we assume $(\Gamma^-_{ir})_{\bar p,\bar p'}=\Gamma^-_{ir}(\bar q)~\forall~p,p'$ \cite{note:gamma}. 

Let us now put  $q_0=0$ (and suppress the $q_0$-argument  from here on)  and  consider a $q$-independent short-ranged (screened) interaction potential, 
yielding $\Gamma^-_{ir}(\bar q)=-U$. 
This allows us to derive from Eq. (\ref{eq:chis})
a simple formula for $\partial\chi_s /{\partial q}$ given by
\beq\label{eq:deltachis}
\frac{\partial\chi_s}{\partial q}(q)
=\frac{\partial\Pi(q)}{\partial q}\frac{1}{(1+U\Pi(q))^2},
\eeq
where $\Pi(q)=\sum_{\bar p} R_{\bar p}(q)/L^{D}$. In the $q\to 0$ limit, one can approximate the term 
$\Pi(q)$ in the denominator of Eq. (\ref{eq:deltachis})
by $\chi_0(0)=-N_e$. The resulting factor $1/(1-UN_e)^2$ in Eq.  (\ref{eq:deltachis}) can be interpreted as a type of 
random phase approximation (RPA) for the electron-hole scattering amplitude \cite{wolff:1960}.
The corrections to the polarization bubble $\Pi(q)$ (dominated by the first bubble correction
to the self-energy) have been calculated in second order
in perturbation theory (in U) at small q by Chubukov and Maslov 
\cite{maslov:2003}. 
The result of 
this perturbative approach is
%\beq
$\delta\Pi(q)=\Pi(q)-\Pi(0)\approx -{4q\chi_{s}(0)\Gamma_s^2}/{3\pi k_F}$,
%\eeq
where 
$\Gamma_s\sim-Um^*/4\pi$ 
denotes the backscattering amplitude. 
When $U N_e\ll 1$, we recover from Eq. (\ref{eq:deltachis}) the known result $\delta \chi_s(q)=\delta \Pi(q)$ \cite{maslov:2003}. 
%If we assume $\Gamma_s=O(1)$ (this is an upper bound because 
%$\Gamma_s $ is a small parameter controlling the perturbation theory),
%then ${\partial\chi_s(q)}/{\partial q}=-{4\chi_{s}(0)\Gamma_s^2}/{3\pi k_F}$.
%%=\frac{4}{3\pi}\Gamma_s^2(\pi)\chi_s(0)/k_F$.

Now we are ready to obtain an estimate for the Curie temperature $T_{c}$.
Replacing $\chi_s(0)$ in $\delta \chi_s(q)$ by its non-interacting limit $\chi_0(0)$, 
and assuming $\Gamma_s=O(1)$ (this is an upper bound because 
$\Gamma_s $ is a small parameter controlling the perturbation theory),
we obtain then from Eq. (\ref{eq:tcsw}) 
 $T_c^{}\sim 25~\mu K$ for  typical 2DEG parameters.
This value of $T_c$ becomes further enhanced by a numerical factor 
(e.g. of order $5$ for $r_s\sim 8$ \cite{GV}) 
%where $r_s$ denotes the dimensionless interaction parameter) 
if one uses an effective renormalized value for the spin susceptibility
 $\chi_S=\chi_s(0)$ 
%(which takes into account some interaction effects) 
instead of $\chi_0(0)$. 
Though $T_c^{}$ is still rather small, it is now finite,
confirming our arguments 
related  to the Mermin-Wagner theorem
that e-e interactions
increase the Curie temperature. 
When $UN_e$ is no longer negligible compared to $1$, $T_c$ is even further enhanced by an additional numerical factor $1/(1-UN_e)^2$ (see Eq. (\ref{eq:deltachis})).
Close to the  ferromagnetic Stoner instability of the electron system, reached when $UN_e\sim 1$,
the Curie temperature $T_c$ for the nuclear system is dramatically enhanced as could have been anticipated.

In the preceeding paragraphs, we replaced $\Gamma_{ir}^-(q)$ by a $q-$independent constant operator. 
One can use instead another approximation called the local field factor approximation (LFFA). 
The idea of the LFFA is to replace the average electrostatic potential by a local field potential
seen by an electron with spin $\s$ (see \cite{GV} for a review).
In this scheme $(\Gamma^-_{ir}(q))_{pp'}\approx -V(q)G_-(q)$, 
where $G_-(q)$ is a local field factor and ${ V}(q)=2\pi
 e^2/\kappa q$ the bare {\em unscreened} Coulomb interaction 
($\kappa$ is the dielectric constant).
%Note that $V$ is the bare unscreened Coulomb interaction as opposed to $U$ 
%which %was a screened interaction. 
Within this approximation scheme
the static spin susceptibility $\chi_s$ becomes
\beq\label{eq:chi_lfft}
\chi_s(q)=\frac{\chi_0(q)}{1+{V}(q)G_-(q)\chi_0(q)}.
\eeq
Determining precisely $G_-(q)$ for all $q$ is still an open issue. However, the asymptotic regimes
are quite well established nowadays \cite{GV}.
 A semi-phenomenological interpolation formula  based on the original Hubbard local field factor 
\cite{hubbard}
and modified in such a way that the compressibility sum rule is exactly satisfied reads \cite{GV,pines}:
\beq\label{eq:lfft}
G_-(q)\approx g_0\frac{q}{q+g_0(1-\chi_P/\chi_S)^{-1}\kappa_2},
\eeq
where $g_0$ is related to the probability of finding two electrons (of opposite spins) at the same position in the electron liquid, $(g\mu_B)^{-2}\chi_P$ is the Pauli susceptibility and $\mu_B$ the Bohr magneton. For non-interacting electrons $\chi_P/\chi_S=1$.
An approximate form for $g_0$ giving good agreement with quantum Monte Carlo (QMC) calculations has been proposed recently by Gori-Giorgi {\it et al.} \cite{gori:2004}: $g_0(r_s)\approx (1+Ar_s+B r_s^2+Cr_s^3)e^{-Dr_s}/2$.
% where $r_s$ denotes the dimensionless density parameter. 
In a 2DEG, $r_s=1/\sqrt{\pi n_e}a_B^*$ where $a_B^*=\kappa/m^*e^2$ is the effective Bohr radius.  The parameters $A=0.088,~B=0.258,~C=0.00037,~D=1.46$ are fitting parameters reproducing QMC results for the 2DEG \cite{gori:2004}.
%Contrary to the previous approximation, $G_-(q){V}(q)$ is now well defined at small $q$.
From Eqs. (\ref{eq:tcsw}) and (\ref{eq:chi_lfft}), one can easily determine $T_{c}$ 
within the LLFA scheme to be given by
\beq \label{eq:tcsw1}
T_{c}^{}=\frac{IA}{2k_B}\sqrt{ \frac{3I}{\pi }}\frac{A}{(\al-1)^2g_0 {V}(a)},
\eeq 
where $\al=(1-\chi_P/\chi_S)^{-1}$ and ${V}(a)$ is the Coulomb potential evaluated at the interatomic distance $a$.
The energy scale $(\al-1)^2g_0 {V}(a)$ can be interpreted as a renormalized screened potential
due to collective interaction effects that are incorporated in the LFFA.
The ratio ${A}/(\al-1)^2g_0 {V}(a)$ can be regarded as the small parameter of our theory.
Quite remarkably, the LFFA  predicts an exponential enhancement of $T_{c}^{}$ 
with increasing interaction parameter $r_s$. For a value of $r_s\sim 5$, this theory already predicts a 
large $T_{c}^{}\sim 25~mK$, a temperature which is routinely achieved nowadays.
Obviously, for some value of $r_s$, the dimensionless parameter ${A}/(\al-1)^2g_0 {V}(a)$
exceeds unity. The truncation of the Schrieffer-Wolff 
transformation at lowest order becomes unjustified
and feedback effects between the electron gas and the nuclear spins, not incorporated in our theory,  become important. Nevertheless for relatively large values of $r_s\lesssim 6$, the condition $A\ll (\al-1)^2g_0{ V}(a)$ is satisfied. 

%Though the LFFA may overestimate $T_c$, the trend in all
Although the spin wave analysis may overestimate $T_c$, the trend in all
the approximation schemes we used is that e-e interactions increase dramatically the Curie 
temperature, possibly into the $mK$ range for large $r_s$ (therefore three orders of magnitude larger than $E_{dip}$ which justifies our starting Hamiltonian). 
We note that the non-perturbative LFFA theory predicts higher $T_c$'s than the perturbative calculation in the short-ranged interaction.
%This prediction urges for an accurate
%prediction of the low $q$ behavior of the exact spin susceptibility in $2D$ which determines $T_c$
%according to Eq. (\ref{eq:tcsw}).
%When our small parameter is close to $1$, the Curie temperature is of order $A$
%which means $T_c=O(1~K)$. This is certainly a far upper limit.
Finally, below $T_c$, the nuclear spins within the 2DEG polarize and generate
an effective magnetic field of order of a few Tesla. This will create a small Zeeman splitting \cite{note:self}
in the 2DEG which should be detectable with e.g. optical or transport methods.

In summary, we have analyzed the Curie temperature $T_c$ of nuclear spins in an interacting 2DEG using a mean field and a spin wave analysis. We have shown that electron-electron interactions considerably enhance the temperature for a ferromagnetic phase transition in the nuclear system,
with $T_{c}$ in the milli-Kelvin range for 2DEGs with $r_s \sim  5-10$.
%, giving values in the milli-Kelvin range for 2DEGs with $r_s \sim  5-10$. 
We thank B. Coish, L. Glazman, L. Kouwenhoven, and A. Yacoby for useful
discussions. This work is supported by the Swiss NSF, NCCR Nanoscience,
ONR, and JST ICORP.


\begin{thebibliography}{30}
\bibitem{loss:1998} D. Loss and D. P. DiVincenzo, Phys. Rev. A {\bf 57}, 120 (1998).
 \bibitem{kroutvar:2004a}
M. Kroutvar, 
%et al.,
Y. Ducommun, D. Heiss, M. Bichler, D. Schuh, G. Abstreiter, and
 J.~J. Finley, 
Nature {\bf 432},  81  (2004).
\bibitem{elzerman:2004} J. M. Elzerman, R. Hanson, L. H. W. van Beveren, B. Witkamp, L. M. Vandersypen, and L. P. Kouwenhoven, Nature {\bf 430}, 431 (2004).

\bibitem{amasha:2006a}
S. Amasha, K. MacLean, I. Radu, D. Zumbuhl, M. Kastner, M. Hanson, and A.
  Gossard, arXiv:cond-mat/0607110  (2006).


\bibitem{petta:2005} J. R. Petta et al., Science {\bf 309}, 2180 (2005).
\bibitem{koppens:2006} F. H. L. Koppens et al., Nature {\bf 442}, 766 (2006).


\bibitem{kikkawa:1998} J. M. Kikkawa and D. D. Awschalom, Phys. Rev. Lett. {\bf 80}, 4313 (1998).

\bibitem{burkard:1999a} G. Burkard, D. Loss, and D. P. DiVincenzo, Phys. Rev. B {\bf 59}, 2070 (1999).
\bibitem{coish:2004a} W. A. Coish and D. Loss, Phys. Rev. B {\bf 70}, 195340 (2004).
\bibitem{khaetskii:2003} A. V. Khaetskii, D. Loss, and L. Glazman, Phys. Rev. Lett. {\bf 88}, 186802 (2002); Phys. Rev. B {\bf 67}, 195329 (2003).  
\bibitem{imamoglu:2003} A. Imamoglu, E. Knill, L. Tian, and P. Zoller, Phys. Rev. Lett. {\bf 91}, 017402 (2003).
\bibitem{bracker:2005a} A. S. Brackner {\it et al.}, Phys. Rev. Lett. {\bf 94}, 047402 (2005).
\bibitem{FN:1940} H. Fr\"ohlich and F. R. N. Nabarro, Proc. Roy. Soc. (London) {\bf A175}, 382 (1940).
%\bibitem{MW} N. D. Mermin, and H. Wagner,  Phys. Rev. Lett. {\bf 17}, 1133 (1966). 
\bibitem{sigrist:1997} H. Tsunetsugu, M. Sigrist, and K. Ueda, Rev. Mod. Phys. {\bf 69}, 809 (1997).
%\bibitem{sigrist:1991} M. Sigrist, H. Tsunetsugu, and K. Ueda, Phys. Rev. Lett. {\bf 67}, 2211 (1991) 
%(we notice that the proof given in this paper is however flawed)
%\bibitem{sigrist:1992}  M. Sigrist, K. Ueda, and  H. Tsunetsugu, Phys. Rev. B {\bf 46}, 175 (1992).

\bibitem{belitz:1997} D. Belitz, T. R. Kirkpatrick, and T. Vojta, Phys. Rev. B {\bf 55}, 9452 (1997).
\bibitem{maslov:2003}  A. V. Chubukov, and D. L. Maslov, Phys. Rev. B {\bf 68}, 155113 (2003).
\bibitem{maslov:2006}
S. Gangadharaiah, D. L. Maslov, A. V. Chubukov, and L. I. Glazman, Phys. Rev. Lett. {\bf 94}, 156407 (2005).
%A. V. Chubukov, D. Maslov, S. Gangadharaiah, and L. I. Glazman, Phys. Rev. B {\bf 71}, 205112 (2005).D. L. Maslov, A. V. Chubukov, and R. Saha, cond-mat/0609102.
\bibitem{efetov:2006} 
%I. L. Aleiner and K. B. Efetov, Phys. Rev. B {\bf 74}, 075102 (2006).
G. Schwiete and K. B. Efetov, cond-mat/0606389. 


\bibitem{note:3d} Strictly speaking, the electron gas is not $2D$ but has a finite thickness of order $5~nm$ and 
therefore contains several layers of $2D$ nuclear spin planes. However, the electronic wave function
is $2D$ and confined in the third dimension. Had we started with a $3D$ lattice of nuclear spins 
where a lattice site is labeled by 
$\vec r_j=(\vec r_{//j},z_j)$, one could reduce the problem to an effective $2D$ one by integrating
out the electronic wave function in the transverse $z$ direction and defining
a new nuclear spin $\vec I_{j//}=\int dz_j |\psi(z_j)|^2\vec I_j$, where $\vec I_j=\vec I(\vec r_j)$
and  $|\psi(z_j)|^2$ is the electron probability density in the transverse direction (with
$\int dz |\psi(z)|^2=1$). In a spin wave mean field treatment of the nuclear spin ferromagnet $I_{j//}=I_j$.

\bibitem{lee:1986} P. A. Lee, T. M. Rice, J. W. Serene, L. J. Sham, and J. W. Wilkins, Comments Condens. Matter Phys. {\bf 12}, 99 (1986).
\bibitem{fs} J. K\"onig, H.-H. Lin, and A. Mc Donald, Phys. Rev. Lett {\bf 84}, 5628 (2000); A. Chattopadhyay, S. Das Sarma, and A. J. Millis, Phys. Rev. Lett. {\bf 87}, 227202 (2001); C. Santos and W. Nolting, Phys. Rev. B {\bf 65}, 144419 (2002).  
\bibitem{GV} G. F. Giuliani and G. Vignale, {\it Quantum Theory of the Electron Liquid},
Cambrigde University Press (2005).

\bibitem{dispersion}
Such a linear spin wave dispersion  is usually associated
with antiferromagnets while  a quadratic one with ferromagnets.
This unusal behavior obtained here is a direct consequence of  e-e interactions giving
a non-zero $\partial\omega_{q} /{\partial q} \sim \partial\chi_s /{\partial q}$ at $q=0$.

\bibitem{Tyablikov} N.N. Boglyubov and S.V. Tyablikov, Dok. Akad. Nauk SSSR {\bf 126}, 53 (1959) [Sov. Phys. Dokl. {\bf 4}, 589 (1959)].
\bibitem{bruno:2001} P. Bruno, Phys. Rev. Lett. {\bf 87}, 137203 (2001).
\bibitem{note:gamma} In  lowest order in the interaction $V$, we have $(\Gamma^-_{ir})_{\bar p\bar p'}=-V_{|\vec p-\vec p'|}$.
\bibitem{wolff:1960} P. A. Wolff, Phys. Rev. {\bf 120}, 814 (1960).
\bibitem{hubbard} J. Hubbard, Proc. Roy. Soc. (London) {\bf A243}, 336 (1957).
\bibitem{pines} N. Iwamoto, E. Krotscheck, and D. Pines,
Phys. Rev. B {\bf 29}, 3936 (1984).
\bibitem{gori:2004} P. Gori-Giorgi, S. Moroni, and G. B. Bachelet
Phys. Rev. B {\bf 70}, 115102 (2004).   
\bibitem{note:self} In order to improve our treatment, one may also use a self-consistent calculation
to determine $T_c$.
In the limit $A\ll g_0{V}(a)$, we expect $T_c$ to be modified by a numerical factor of order $1$.
%\bibitem{bruno:2001} P. Bruno, Phys. Rev. Lett. {\bf 87}, 137203 (2001).
%\bibitem{spintronics} {\it Semiconductors Spintronics and Quantum Computation} edited by D. D. Awschalom, D. Loss and N. Samarth  (Springer,Berlin, 2002).
\end{thebibliography}
\end{document}